\documentclass[12pt]{revtex4}
\usepackage{amsmath}

\begin{document}

\title{Spin-Electromagnetic Hydrodynamics 
and Magnetization Induced by Spin-Magnetic Interaction}

\author{T. Koide}
\affiliation{Instituto de F\'{\i}sica, Universidade Federal do Rio de Janeiro, C.P.
68528, 21941-972, Rio de Janeiro, Brazil}

\begin{abstract}
The hydrodynamic model including the spin degree of freedom and the electromagnetic field was discussed.
In this derivation, we applied electromagnetism for macroscopic medium proposed by Minkowski.
For the equation of motion of spin, we assumed that the hydrodynamic representation of the Pauli equation 
is reproduced when the many-body effect is neglected. 
Then the spin-magnetic interaction in the Pauli equation was converted to a part of the magnetization.
The fluid and spin stress tensors induced by the many-body effect were obtained by employing the algebraic positivity of the entropy production 
in the framework of the linear irreversible thermodynamics, including 
the mixing effect of the irreversible currents.
We further constructed the constitutive equation of the polarization and the magnetization. 
Our polarization equation is more reasonable compared to another result obtained 
using electromagnetism for macroscopic medium proposed by de Groot-Mazur.
\end{abstract}

\maketitle

\section{Introduction}

Hydrodynamics has been used as an effective model for discussing many-body effects and collective motions not only in 
classical systems but also in quantum ones. 
In the physics of relativistic heavy-ion collisions,  
it has been confirmed experimentally that the classical relativistic hydrodynamic model can explain the experimental behaviors of collective 
motions of microscopic quantum particles qualitatively \cite{rev_hyd}.
In this case, the difference of quantum statistics for constituent particles of the fluid, 
such as boson and fermion, does not modify the structure of hydrodynamics. 
That is, it is considered that the electron fluid and the pion fluid obey the essentially same 
hydrodynamic model except for the equation of states.
This idea is, however, not trivial under the existence of the magnetic field 
because of the spin-magnetic interaction for fermions.
Then the spin degree of freedom can play an important role even in constructing a hydrodynamic 
model coupling with the electromagnetic field.

In fact, the spin effect under the electromagnetic field has been attracted various attentions in 
the physics of relativistic heavy-ion collisions \cite{cme}.
There, the possibility of the formation of the coherent magnetic field is discussed 
and it is expected that the effect of this magnetic field to the spin degree of freedom 
can be observed through experiments.
To investigate this possibility more dynamically, we need to construct a relativistic hydrodynamic model including the 
spin degree of freedom.
However, such a model has not yet been established even in the non-relativistic case.

For example, the spin effect is important also in the physics of quantum plasma 
and a magnetohydrodynamics with the spin degree of freedom was proposed recently \cite{mark}. 
In this model, the Pauli equation is re-expressed using classical variables such as 
the particle probability density, 
the particle velocity and the spin vector. 
By assuming that each particle of a quantum plasma follows the same one-particle Pauli equation, 
the hydrodynamic equation is obtained by the independent sum of these constituent particles of the plasma, 
neglecting the effect of the off-diagonal quantum correlations.
It is worth mentioning that 
the same re-expression of the Pauli equation with the classical hydrodynamic variables 
was already done by several groups in 1950s \cite{taka,hol}.

However, it is not possible to apply this result directly to the physics of 
the relativistic heavy-ion collisions, even if we ignore the difference between 
relativistic and non-relativistic models.
One of the reasons is that the barotropic fluid is considered 
and hence the equation for the energy density is not derived in Ref. \cite{mark}. 
Furthermore, the positivity of the entropy production is not used in the derivation
and hence the viscous tensors and the heat current are not obtained.
Thus we need to apply the linear irreversible thermodynamics (LIT) for constructing 
a hydrodynamic model with spin.

Another problem is that the definition of the electromagnetic momentum density in macroscopic medium 
has not yet been established \cite{rev,eu-oppen,fk}.
In fact, there are experimental attempts to determine these definitions 
but no conclusive result is obtained because of the difficulty of the division of 
the electromagnetic part and the material part \cite{rev}.

Theoretically, there are at least three different proposals: 
the definitions of Minkowski \cite{mink}, Hertz-Abraham \cite{abra} and de Groot-Mazur \cite{degroot}.
In the Minkowski theory, the electromagnetic momentum density is defined by ${\bf D} \times {\bf B}$ (for the definitions of variables, see Sec. III).
However the form of the Maxwell stress tensor is asymmetric in this theory (see Eq. (\ref{em_t})). 
As is mentioned in Sec. IV B, the symmetry of this tensor is related to the angular momentum conservation.
To introduce the symmetric Maxwell stress tensor, Abraham employed $\varepsilon_0 \mu_0 {\bf E} \times {\bf H}$ as the 
definition of the electromagnetic momentum density, following the lead of Hertz.
de Groot, Mazur and Suttorp discussed this problem in two different ways, LIT and a certain microscopic model, 
and finally reproduced Hertz-Abraham's momentum density.
However, their Maxwell stress tensor has the antisymmetric form.
See Refs. \cite{rev,eu-oppen} for details.
The derived hydrodynamic models depend on the choice of these definitions. 
In Ref. \cite{eu-oppen}, Eu and Oppenheim derived 
the three hydrodynamic models using these three different definitions 
and concluded that the definition of Minkowski is promising. 
However, as was claimed in Ref. \cite{fk}, their 
discussion is incomplete because they could not show the positivity of the entropy production.

On the other hand, Felderhof and Kroh employed the definition of de Groot-Mazur and succeeded 
in deriving the model of electromagnetic hydrodynamics which is consistent with the positivity of the entropy production 
at the last step of the derivation \cite{fk}.
They introduced the internal angular momentum as another hydrodynamic variable, which may be interpreted as spin.
If we cannot construct spin-hydrodynamic models consistent with LIT by using other definitions, 
the definition of de Groot-Mazur will be preferred, as was concluded in Ref. \cite{fk}.
To confirm this conclusion, it is worth investigating once again the derivation with other definitions.

In this paper, we derive the spin-electromagnetic hydrodynamics (sEMHD) 
by combining the above two approaches.
First, we define the equation of motion of the spin vector 
so as to be consistent with the hydrodynamic representation of the Pauli equation.
Then the spin-magnetic interaction in the Pauli equation induces the magnetization.
On the other hand, the many-body effects of fluids are taken 
into account through the modification of the fluid and spin stress tensors which are obtained 
by employing the algebraic positivity of the entropy production in the framework of LIT.

As the definitions of the momentum density, we apply the definition 
of Minkowski.
The electromagnetic hydrodynamics with the definition of Minkowski was already discussed 
in Ref. \cite{imai}.
We incorporate the information from the Pauli equation to this result.
We further derive the explicit forms of the irreversible currents taking into account the Curie principle and 
the relaxation equations of the polarization and the magnetization, which were not discussed in Ref. \cite{imai}.

For the purpose of the physics of relativistic heavy-ion collisions, 
we should discuss a relativistic model. However, as was discussed above, 
there are various unsolved problems even in the non-relativistic case. 
Thus we focus on the derivation of the non-relativistic sEMHD in the present work. 
The consistent relativistic sEMHD should reproduce the result of this paper 
in the non-relativistic limit.

This paper is organized as follows.
In Sec. II, we summarize the hydrodynamic expression of the Pauli equation following Refs. \cite{taka,hol}.
In Sec. III, the Maxwell equation for macroscopic medium is introduced following the definition of Minkowski. 
Then we consider the magnetic polarization induced through the spin-magnetic interaction in the Pauli equation.
In Sec. IV, we extend the result of Sec. II to hydrodynamics, so as to conserve 
the momentum and the angular momentum.
To determine the stress tensors of fluid and spin, we employ the algebraic 
positivity of the entropy production, and the equation of sEMHD is obtained in Sec. V. 
The result depends on the definition 
of the thermodynamic pressure. 
By using this fact, it is shown that our result is consistent with that by Korteweg-Helmholtz 
in Sec. VI.
The relaxation equations of the polarization and the magnetization are derived in Sec. VII.
The comparison with other hydrodynamic models is discussed in Sec. VIII. 
Section IX is devoted to concluding remarks.

\section{Hydrodynamic expression of Pauli equation} \label{sec:pauli}

Let us consider the Pauli equation coupling with external gauge potentials ${\phi}$ and ${\bf A}$, 
\begin{equation}
i\hbar \frac{\partial}{\partial t} \psi 
= 
\left[ \frac{1}{2m} \left( \frac{\hbar}{i}\nabla - \frac{e}{c}{\bf A} \right)^2 + e \phi 
- \frac{\gamma_g}{2} \frac{e\hbar}{2mc}{\bf \sigma}\cdot {\bf B} \right]
\psi,
\end{equation}
where $c$, $e$, $m$, ${\bf \sigma}$ and $\gamma_g$ are the speed of light, 
the electric charge, the mass of a particle, the Pauli matrix and the g factor, respectively.
The magnetic flux density ${\bf B}$ is expressed by using 
the vector potential ${\bf A}$ as ${\bf B} = {\rm rot }{\bf A}$.  
The wave function $\psi$ has two components,
\begin{eqnarray}
\psi = 
\left(
\begin{array}{c}
\psi_1 \\
\psi_2
\end{array}
\right),
\end{eqnarray}

Bohm et al. \cite{taka} extended the idea by Madelung \cite{mad,wong} and obtained the hydrodynamic form 
of the Pauli equation.
For this purpose, let us introduce the following three variables,
\begin{subequations}
\begin{eqnarray}
\rho &=& \psi^\dagger \psi, \\
{\bf v} &=& \frac{\hbar}{2mi}\frac{\psi^\dagger \nabla \psi - \psi \nabla \psi^\dagger}{\rho} - \frac{e}{mc} {\bf A}, \\
{\bf s} &=& \frac{\hbar}{2} \frac{\psi^\dagger {\bf \sigma} \psi}{\rho}. \label{spinvec}
\end{eqnarray}
\end{subequations}
The first two variables denote the particle probability density and the probability velocity, respectively.
The last vector ${\bf s}$ represents the spin degree of freedom.
One can easily confirm that the magnitude of this spin vector is normalized as ${\bf s}^2 = \hbar^2/4$.

Then all the quantities calculated from the Pauli equation can be reproduced by the 
combinations of the above three c-number variables \cite{taka,hol}. 
In fact, the evolution equations of the three variables are expressed in the closed form.
Directly from the Pauli equation, these equations are given by 
\begin{subequations}
\begin{eqnarray}
\partial_t \rho 
&=&
- \frac{\hbar}{2im} \sum_k \partial_k \cdot \{\psi^\dagger 
(  \partial_k  - \overleftarrow{\partial}_k ) \psi \}
+ 
\frac{e}{mc} \nabla \cdot ( \rho {\bf A}  ) , \\
\partial_t (\rho v^{i})
&=& 
\frac{\hbar^2}{4m^2} \sum_k 
\partial_k ( \psi^\dagger \overleftrightarrow{\partial_i}\overleftrightarrow{\partial_k} \psi )
+ \frac{e}{c m} 
\left[
\sum_k   A^k \partial_k \left\{
\left( 
v^i + \frac{e}{mc}A^i
\right) \rho  \right\}
\right. \nonumber \\
&& \left. 
+  \sum_k (\partial_k A^k) 
\left\{ \rho
\left( 
v^i + \frac{e}{mc}A^i
\right)  \right\}
+ \sum_k ( \rho v^k \partial_{i} A^k)
 \right] \nonumber \\
&&  
+ \frac{e}{m} E^i \rho 
+ \frac{e\gamma_g }{2m^2 c} \sum_k \rho s^k (\partial_{i} B^k)  
- \frac{e}{mc} A^{i} \partial_t \rho ,\\
\partial_t (\rho s^i)
&=&
- \frac{\hbar^2}{4 m i} \sum_k \partial_k \left[   
\psi^\dagger \sigma^{i} \partial_k \psi
-
(\partial_k \psi^\dagger) \sigma^{i} \psi
\right]
+ \frac{e}{m c} \sum_k \partial_k (A^k \rho {\bf s}^{i})
 + \frac{e\gamma_g}{2mc} \rho [ {\bf s} \times {\bf B} ]_i ,
\end{eqnarray}
\end{subequations}
where 
$\overleftrightarrow{\partial_i} = \partial_i - \overleftarrow{\partial}_i$. 
There are still terms which 
are expressed with the wave function $\psi$ on the right hand sides of the equations.
To eliminate this dependence, we use the following relations, 
\begin{subequations}
\begin{eqnarray}
&& \psi^* \overleftrightarrow{\partial}_i \overleftrightarrow{\partial}_k \psi
= 
\left\{ -4 \frac{m^2}{\hbar^2}\rho \left( v + \frac{e}{mc}A\right)^i \left( v + \frac{e}{mc}A\right)^k 
+ \rho \partial_i \partial_k \ln \rho
-
\frac{4}{\hbar^2}\rho \sum_l
(\partial_i  s_l) (\partial_k s_l) \right\}, \nonumber \\
\\
&& i \left\{ (\psi^* \sigma_k \partial_l \psi) - (\psi \sigma_k \partial_l \psi^*)  
  \right\} 
=
\frac{4}{\hbar^2} \rho [ {\bf s} \times  \partial_l {\bf s}]_k 
- s^k m \rho \left( v^l + \frac{e}{mc}A^l\right).
\end{eqnarray}
\end{subequations}
To derive these expressions, we used 
\begin{subequations}
\begin{eqnarray}
\psi^* \partial_i \partial_k \psi + (\partial_i \psi^*)(\partial_k \psi) 
&=& \frac{1}{2} \left( \partial_i \partial_k \rho + \frac{2mi}{\hbar}\partial_i 
\left\{ \rho \left( v + \frac{e}{mc}A\right)^k \right\} \right), \\
\frac{4}{\hbar^2}\rho (s_i \partial_l s_j - s_j \partial_l s_i) 
&=& \sum_k
i \epsilon_{ijk} \left\{ (\psi^* \sigma_k \partial_l \psi) - (\psi \sigma_k \partial_l \psi^*)  
- i m \rho s^k \left( v^l + \frac{e}{mc}A^l\right)  \right\}. \nonumber \\ 
\end{eqnarray}
\end{subequations}

Then the hydrodynamic expression of the Pauli equation is finally given by
\begin{subequations}
\begin{eqnarray}
\frac{D}{Dt} \rho 
&=&
-  \rho \sum_k \partial_k v^{k} , \\
\frac{D}{Dt} v^{i}
&=&   
\frac{1}{\rho} \sum_k \partial_k \left\{ 
\frac{\hbar^2}{4m^2}  \rho \partial_i \partial_k \ln \rho
-\frac{1}{m^2}  
\rho \sum_l 
(\partial_i s_l) (\partial_k s_l) \right\} \nonumber \\
&& + \frac{e\gamma_g}{2m^2c} \sum_k  s^k \partial_{i} B^k    
+ \frac{e}{m} \left[{\bf E} + \frac{1}{c}{\bf v} \times {\bf B} \right]_i , \label{vel}\\
\frac{D}{Dt} {\bf s}^i
&=&
 \frac{1}{m \rho} \sum_k 
\partial_k 
\left\{
\rho [ {\bf s} \times  \partial_k {\bf s}]_i 
\right\}
 + \frac{e \gamma_g}{2mc} [ {\bf s} \times {\bf B} ]_i ,  \label{spin} 
\end{eqnarray}
\end{subequations}
where we introduced the material (Lagrange) derivative as
\begin{equation}
\frac{D}{Dt} = \partial_t + {\bf v}\cdot \nabla .
\end{equation}

The first equation represents the conservation of the particle probability.
The second one is the equation of the probability velocity and 
is sometimes considered as the Euler equation by interpreting the 
first term on the right hand side as the pressure gradient \cite{taka,hol}.
The last equation gives the evolution of the spin vector.
That is, the above equations can be regarded as the ideal fluid model with spin and the external electromagnetic field.

We will use this result to construct the evolution equation of spin in the hydrodynamic model.
The above result is exact as the quantum dynamics of one particle, but we assume that the same equation is 
approximately applicable even to the non-interacting many-body systems.
Then the particle probability density $\rho$ is interpreted as the particle number density and 
the probability velocity ${\bf v}$ as the fluid velocity.

\section{Maxwell equation for macroscopic medium with spin-magnetic interaction}

Our model will be constructed so as to satisfy the total energy, momentum and angular momentum conservations \cite{imai}. 
Here {\it total} means the sum of the contributions from the fluid and the electromagnetic field.
For this, we have to define the energy and momentum densities of the electromagnetic field 
for macroscopic medium. 
In the present work, we adapt the Minkowski definition.

Differently from the case of electromagnetism in vacuum, the electric and magnetic flux densities, 
${\bf D}$ and ${\bf B}$, 
are not parallel to the electric and magnetic fields, ${\bf E}$ and ${\bf H}$, and given by 
\begin{subequations}
\begin{eqnarray}
{\bf D} &=& \varepsilon_0 {\bf E} + {\bf P}, \\
{\bf B} &=& \mu_0 {\bf H} + {\bf M} + {\bf M}_{Pauli},
\end{eqnarray}
\end{subequations}
where $\varepsilon_0$, $\mu_0$ and ${\bf P}$ are the dielectric constant, the magnetic permeability
and the polarization, respectively.
Here we consider two magnetizations. 
The quantity ${\bf M}_{Pauli}$ denotes the magnetization induced by the spin-magnetic interaction 
in the Pauli equation.
This contribution is obtained so as to satisfy the momentum and angular momentum conservations as we will see soon later. 
On the other hand, ${\bf M}$ represents the possible higher-order spin contributions neglected in the Pauli equation \cite{Itz-Zu}.
If we assume that the magnetization is induced only through the spin-magnetic interaction in the Pauli equation, 
${\bf M}$ disappears.
In this paper, this term is obtained by employing the positivity of the entropy production. 
See Sec. \ref{chp:pola}.

As the Maxwell equation for macroscopic medium, we adapt the following equations,
\begin{subequations}
\begin{eqnarray}
\nabla \cdot {\bf D} &=& \rho_e , \\
\nabla \cdot {\bf B} &=& 0 , \\
\nabla \times {\bf E} + \frac{\partial {\bf B}}{\partial t} &=& 0, \\
\nabla \times {\bf H} - \frac{\partial {\bf D}}{\partial t} &=& {\bf J},
\end{eqnarray}
\end{subequations}
where ${\bf J}$ is the electric current density, which 
is decomposed into the two parts: one is the convection current $\rho_e {\bf v}$, and 
the other the conduction current ${\bf j}$,   
\begin{eqnarray}
{\bf J} = \rho_e {\bf v} + {\bf j}.
\end{eqnarray}
See also the discussion of chapter XIII in Ref. \cite{degroot}.

Following the definitions of Minkowski, the energy and momentum conservations are expressed by 
\begin{eqnarray}
&& \partial_t U_{em} + {\rm div}~{\bf S} = - W, \label{em-ener}\\
&& \partial_t g^{i} =  \partial_k T^{ik} - f^{i}_{em}. \label{em-mom}
\end{eqnarray}
Here we introduced, respectively, the electromagnetic energy density $U_{em}$, 
the Poynting vector ${\bf S}$, the electromagnetic momentum density ${\bf g}$ 
and the Maxwell stress tensor $T^{ik}$ as
\begin{subequations}
\begin{eqnarray}
U_{em} &=& \frac{1}{2} ({\bf D}\cdot {\bf E} + {\bf B} \cdot {\bf H}), \label{uem} \\
{\bf S} 
&=& {\bf E} \times {\bf H}, \\
{\bf g} &=& {\bf D} \times {\bf B}, \\
T^{ik} 
&=& E^{i} D^{k} + H^{i} B^{k} - U_{em} \delta^{ik}. \label{em_t}
\end{eqnarray}
\end{subequations}
The source term of the energy density and the Lorentz force in macroscopic medium are given by
\begin{eqnarray}
W 
&=& {\bf J}\cdot {\bf E} + {\bf E}\cdot \partial_t {\bf D} + {\bf H}\cdot \partial_t {\bf B} - \partial_t U_{em}, \\
f^i_{em} 
&=& 
\rho_e E^{i} + [ {\bf J} \times {\bf B} ]^{i}
+ \sum_k (D^k \partial_i E^k + B^k \partial_i H^k) - \partial_i U_{em}, \label{fem1}
\end{eqnarray}
respectively.

\section{Spin-electromagnetic hydrodynamics}

To obtain the evolution equation of fluids, we have to extend the result of the Pauli equation to the many-body case.
For this purpose, we introduce new variables: the stress tensor of the fluid $p^{ij}$ and the stress tensor of spin $C^{ij}$.
These stress tensors are, first, fixed by comparing with the results in Sec. \ref{sec:pauli}, and the deviations  
induced by the many-body effects are obtained by applying LIT in the next section.

For the discussion of hydrodynamics, it is more convenient to use 
the mass density defined by $\rho_m = m \rho$ where $m$ is 
the mass of the constituent particle of our fluid.
For the mass equation, we adapt the usual conservation equation,
\begin{equation}
\frac{D}{Dt} \rho_m = - \rho_m \nabla \cdot {\bf v},
\end{equation}
and, correspondingly, we introduce a new variable to represent the spin vector,
\begin{equation}
{\bf s}_m = \frac{\bf s}{m}.
\end{equation}
Then the total spin of the fluid is given by $\int d^3 {\bf x} \rho_m {\bf s}_m$.

Of course, in the case of many-particle systems such as hydrodynamics, 
${\bf s}_m$ does not contain the complete information of spin in a many-body quantum system 
and our spin-hydrodynamic model is an approximation which 
will be justified when the contributions from the off-diagonal components of spin 
(interference of different spins) are absorbed into the 
many-body effect of the spin stress tensor which will be obtained by LIT.
The same off-diagonal part is neglected even in the derivation of the spin-magnetohydrodynamics in Ref. \cite{mark}.
As an indirect example to justify this picture, see the discussion in Sec. 9.6 of Ref. \cite{hol}, 
where the exact hydrodynamic expression of the two electrons system 
is discussed and the off-diagonal parts affect only a part of the spin stress tensor.
In our approach, we assume that this off-diagonal parts is taken into account through the 
modification of the stress tensors obtained by using LIT.

\subsection{Momentum conservation}

We consider an arbitrary volume $V$ whose surface is given by $S$. The normal vector of the surface 
is given by ${\bf n}$. 
The total momentum in this region is given by the sum of the momenta of the fluid and of 
the electromagnetic field.
Then the general expression of the momentum conservation is expressed as
\begin{eqnarray}
\partial_t \int_V dV (\rho_m v^{i}+g^{i} ) = -  \sum_k \int_S dS \rho_m v^{i}  v^{k} n^k +  \sum_k \int_S dS p^{ik} n^k 
+  \sum_k \int_S dS T^{ik}n^k + \int_V dV \rho_m K^{i} . \nonumber \\
\end{eqnarray}
Here, for the sake of generality, we introduced an external force $K^i$ which violates the momentum conservation.
This should be satisfied for any volume and then we obtain the following equation,
\begin{eqnarray}
\rho_m\frac{D}{Dt} v^{i} = \sum_j \partial_j p^{ij} + \rho_m K^{i} + f^{i}_{em}. 
\end{eqnarray}

To reproduce Eq. (\ref{vel}), we identify 
\begin{subequations}
\begin{eqnarray}
 K^i 
&=& 0, \\
p^{ij} 
&=&
(p^s)^{ij} + (p^a)^{ij}, \\
(p^s)^{ij} 
&=&
(\tilde{p}^s)^{ij} +  
\frac{\hbar^2}{4m^2}  \rho_m \partial_i \partial_j \ln \rho_m
- \rho_m \sum_l 
(\partial_i s^l_m) (\partial_j s^l_m) + \frac{\gamma_g \rho_e}{2c} \delta^{ij} {\bf s}_m \cdot {\bf B} \label{ps1},
\end{eqnarray}
\end{subequations}
where $(p^s)^{ij}$ and $(p^{a})^{ij}$ are the symmetric and anti-symmetric parts of 
the stress tensor of the fluid for the exchange of the indices $i$ and $j$.
The terms $(\tilde{p}^s)^{ij}$ and $(p^{a})^{ij}$ represent the modifications of the stress tensor 
due to the many-body effect which will be determined later by employing LIT.

The term $\frac{\gamma_g \rho_e}{2c} \sum_k s^k_m \partial_i B^k$ in Eq. (\ref{vel}) is reproduced from 
the magnetization in the Lorentz force. 
In fact, by choosing 
\begin{equation}
{\bf M}_{Pauli} = \mu_0 \frac{\gamma_g \rho_e}{2c} {\bf s}_m , \label{eqn:mpauli}
\end{equation}
the Lorentz force is expressed as 
\begin{eqnarray}
f^{i}_{em} 
= 
\rho_e E^{i} + [ {\bf J} \times {\bf B} ]^{i}
+ \sum_l \left\{ D^l \partial_i E^l + B^l \partial_i \frac{1}{\mu_0}({\bf B} - {\bf M})^l \right\} - \partial_i U_{em} 
- \frac{\gamma_g \rho_e}{2c} \sum_l  B^l \partial_i {\bf s}^l_m. \label{eqn:fm2} \nonumber \\
\end{eqnarray}
Then, by combining the contributions from the last terms of Eqs. (\ref{ps1}) and (\ref{eqn:fm2}), 
the term $\frac{\gamma_g \rho_e}{2c} \sum_k s^k_m \partial_i B^k$ is reproduced.

In short, we can construct the equation consistent with the momentum conservation 
when we choose ${\bf M}_{Pauli}$ by Eq. (\ref{eqn:mpauli}).
The consistency of this ${\bf M}_{Pauli}$ is confirmed 
in the discussion of the next subsection, the angular momentum conservation.

\subsection{Angular momentum conservation} \label{chap:ang}

Usually, in particular in the relativistic case, 
hydrodynamic models are derived so as to be consistent with 
the conservation of energy and momentum but the angular momentum is not considered explicitly, because 
this conservation is automatically satisfied when the energy-momentum tensor is chosen to be symmetric. 
In the present case, however, the Maxwell stress tensor is not symmetric $T^{ij} \neq T^{ji}$ and 
we need to employ the angular momentum conservation for the construction of the consistent model.

The total angular momentum density is given by the sum of the fluid part ${\bf r} \times  \rho_m  {\bf v}$, 
the electromagnetic part ${\bf r} \times {\bf g}$ and the spin part $\rho_m {\bf s}_m$.
Then the conservation for the angular momentum of the total system is expressed as  
\begin{eqnarray}
&& \partial_t \int_V dV [{\bf r} \times ( \rho_m  {\bf v} + {\bf g}) ]^{i} 
+ 
\partial_t \int_V dV \rho_m s^{i}_m \nonumber \\
&=& 
- \int_S dS [ {\bf r} \times \rho_m  {\bf v} 
+ {\bf s}_m
]^{i} ({\bf v}\cdot {\bf n}) 
+ \int_S dS \sum_{jkl} \epsilon_{ijk} r^{j}(p^{kl} n^{l} + T^{kl} n^{l} ) \nonumber \\
&& + \int_S dS \sum_{ij} C^{ij} n^{j} + \int_V dV \rho_m [{\bf r} \times {\bf K}+ {\bf G}]^{i} 
, \label{amom_int}
\end{eqnarray}
where ${\bf G}$, if it exists, means the source of the angular momentum. 
That is, the last term on the right hand side represents 
the violation of the angular momentum conservation.
From this expression, we obtain the following differential equation, 
\begin{eqnarray}
\rho_m \frac{D}{Dt}  s^{i}_m  
&=&
\sum_{jk} \epsilon_{ijk}  p^{kj} + \sum_{j}\partial_j C^{ij}
+ [{\bf D} \times {\bf E} + {\bf B} \times {\bf H}]^{i} + \rho_m G^{i}. \label{dif_ang1}
\end{eqnarray}
To derive this expression, we used the momentum equation obtained in the previous subsection.
Note that the third term on the right hand side of Eq. (\ref{dif_ang1}) disappears in vacuum 
because ${\bf D} \parallel {\bf E}$ and ${\bf B} \parallel {\bf H}$.

Using the definition of ${\bf M}_{Pauli}$ given by Eq. (\ref{eqn:mpauli}), 
Eq. (\ref{spin}) is reproduced by the following identification,  
\begin{subequations}
\begin{eqnarray}
G^i 
&=& 0, \\
C^{ij} 
&=& \rho_m [{\bf s}_m \times \partial_j {\bf s}_m]_i + \tilde{C}^{ij},
\end{eqnarray}
\end{subequations}
where the possible modification of the spin stress tensor by the many-body effect 
is denoted by $\tilde{C}^{ij}$.
The interaction term ${\bf s}_m \times {\bf B}$ 
in Eq. (\ref{spin}) is induced from ${\bf B} \times {\bf H}$ in Eq. (\ref{dif_ang1}) through ${\bf M}_{Pauli}$.

\section{Local thermal equilibrium and energy equation} \label{chap:lte}

In this section, we calculate the modifications of the stress tensors, 
$(\tilde{p}^s)^{ij}$ and $\tilde{C}^{ij}$ by employing LIT.

For this purpose, we first calculate the energy equation.
Suppose that spin can be expressed as ${\bf s}_m = I {\bf \Omega}$ using the moment of inertia $I$ per unit mass and the angular velocity ${\bf \Omega}$ as in the case of the classical angular momentum. 
Then the total energy density is expressed as 
\begin{equation}
\frac{\rho_m}{2}({\bf v}^2 + I {\bf \Omega}^2) + \rho_m U + U_{em}. \label{ener-den-con}
\end{equation}
These four terms represent, respectively, the kinetic, rotational, internal and electromagnetic energy densities. 
Here we introduced the internal energy per unit mass by $U$.
The validity of 
the assumption of the spin rotational energy used here should be argued carefully.
We will discuss this point in the concluding remarks.

The energy conservation is, then, expressed as 
\begin{eqnarray}
&& \partial_t \int_V dV 
\left[ \frac{\rho_m}{2}({\bf v}^2 + I {\bf \Omega}^2) + \rho_m U + U_{em} 
 \right] \nonumber \\
&=&
- \int_S dS \left[ \frac{\rho_m}{2}({\bf v}^2 + I {\bf \Omega}^2) + \rho_m U 
\right] ({\bf v} \cdot {\bf n})
+ \int_S dS \sum_{ij} p^{ij} n^{j} v^{i} \nonumber \\
&& + \int_S dS \sum_{ij} C^{ij} n^j \Omega^{i} 
+ \int_V dV \rho_m ({\bf K}  \cdot {\bf v}) + \int_V dV \rho_m ({\bf G} \cdot {\bf \Omega}) \nonumber \\ 
&& - \int_S dS ({\bf S}\cdot {\bf n}) - \int_V dV (\nabla \cdot {\bf q}), 
\end{eqnarray}
where ${\bf q}$ denotes the heat current. 
The corresponding differential equation is given by
\begin{eqnarray}
\rho_m \frac{D}{Dt}  U
&=& 
\frac{D}{Dt} U_{em} 
+ {\bf j} \cdot [ {\bf E} + ({\bf v} \times {\bf B})]  
- \left[ {\bf D} \cdot \frac{D^*}{Dt} {\bf E} + {\bf B}  \cdot \frac{D^*}{Dt} {\bf H}  \right] \nonumber \\
&& 
- \sum_{ijk}\epsilon_{ijk}p^{kj} \Omega^i 
+  \sum_{ik}p^{ik} \partial_k v^{i} 
+ \sum_{ik} C^{ik}\partial_k \Omega^{i} -\nabla \cdot {\bf q} , \label{eq_ener}
\end{eqnarray}
where we introduced the material derivative with rotation as
\begin{equation}
\frac{D^*}{Dt} {\bf X} = \frac{D}{Dt} {\bf X} - {\bf \Omega} \times {\bf X},
\end{equation}
for an arbitrary vector ${\bf X}$.

In general, $(\tilde{p}^s)^{ij}$ has $6$ degrees of freedom. 
In this work, following the usual phenomenological derivation of the Navier-Stokes equation, 
we parametrize it by the three quantities, the thermodynamic pressure $P$, the bulk viscous pressure $\tau_D$ 
and the shear stress tensor $(\tau^s)^{ij}$,
\begin{eqnarray}
(\tilde{p}^s)^{ij} + \frac{\hbar^2}{4m^2} \rho_m \partial_j \partial_k \ln \rho_m
- \rho_m \sum_l (\partial_j s^l_m)(\partial_k s^l_m) = -P \delta^{ij} +\tau^{D} \delta_{ij} + (\tau^{S})^{ij} ,
\end{eqnarray}
where $(\tau^S)^{ij} = (\tau^S)^{ji}$ and $\sum_{i} (\tau^S)^{ii} = 0$. 
The term $ \frac{\hbar^2}{4m^2} \rho_m \partial_j \partial_k \ln \rho_m
- \rho_m \sum_l (\partial_j s^l_m)(\partial_k s^l_m)$ 
is known as a kind of pressure comes from quantum fluctuation \cite{taka,hol}. 
In the above expression, we define the pressures and the stress tensor so as to renormalize this quantum effect into their definitions.
It is possible to consider the more complex decomposition of the fluid stress tensor. 
See p.311 of Ref. \cite{degroot} for details.

Substituting $p^{ij}$ and $C^{ij}$ into Eq. (\ref{eq_ener}), we obtain the energy equation, 
\begin{eqnarray}
\rho_m \frac{D}{Dt}  U
&=&
-\frac{D}{Dt} U_{em} 
+ 
 {\bf E} \cdot \frac{D^*}{Dt} {\bf D}  
+ {\bf H}  \cdot \frac{D^*}{Dt} {\bf B}
+ P \frac{1}{\rho_m} \frac{D}{Dt} \rho_m  \nonumber \\
&&
+ {\bf j}\cdot [{\bf E} + ({\bf v} \times {\bf B})]  
 + \left[ \tau^D + \frac{\gamma_g \rho_e}{2c} ({\bf s}_m \cdot {\bf B} ) 
\right] \theta  
+ \sum_{jk} (\tau^S)^{jk} \frac{1}{2}e^{jk} 
\nonumber \\ 
&& 
- {\bf p}^a_v \cdot \left( {\bf \Omega} - \frac{1}{4} {\bf \omega}_v \right)
+ \sum_{jk} \tilde{C}^{jk} \partial_k \Omega^{j}
 -\nabla \cdot {\bf q} ,
\end{eqnarray}
where 
\begin{subequations}
\begin{eqnarray}
&& (p^a_v)^i = - \sum_{jk} \epsilon_{ijk} (p^a)^{jk}, \\
&& \omega^{ij} = \partial_j v^{i} - \partial_i v^{j}, \\
&& (\omega_v)^i = - \sum_{jk} \epsilon_{ijk} \omega^{jk},\\
&& e^{ij} = \partial_j v^{i} + \partial_i v^{j} - \frac{2}{3}\theta \delta_{ij}, \\
&& \theta = \sum_i \partial_i v^{i}.
\end{eqnarray}
\end{subequations}

In order to obtain the thermodynamic relation of our system, 
we introduce the following quantities per unit mass,
\begin{subequations}
\begin{eqnarray}
\hat{U}_{em} &=& \frac{U_{em}}{\rho_m}, \\
\hat{\bf D}  &=& \frac{\bf D}{\rho_m}, \\
\hat{\bf B}  &=& \frac{\bf B}{\rho_m}. 
\end{eqnarray}
\end{subequations}
By rewriting the energy equation with these variables, we find that the total energy, which is given by the sum of 
the internal energy per unit mass $U$ and the electromagnetic energy per unit mass $\hat{U}_{em}$, 
can be regarded as a function of $\rho_m$, $\hat{\bf D}$, $\hat{\bf B}$ and additional contributions 
which will be interpreted as heat.
From this behavior, we assume the following thermodynamic relation in this work, 
\begin{eqnarray}
d U + d \hat{U}_{em}
&=&
 {\bf E} \cdot d \hat{\bf D}  
+ {\bf H} \cdot d \hat{\bf B}  
- (P + U_{em} ) 
d \frac{1}{\rho_m} + TdS_{en}, \label{eos1}
\end{eqnarray}
where $T$ and $S_{en}$ are temperature and the entropy per unit mass, respectively.
The similar thermodynamic relation is obtained in Sec. 5 of Ref. \cite{chu}.
This result means that the thermodynamic work is modified from $P d (1/\rho_m)$ 
to $(P + U_{em}) d(1/\rho_m) - {\bf E} \cdot d \hat{\bf D} - {\bf H} \cdot d \hat{\bf B}$. 
This can be shown by calculating the thermodynamic work in the quasi-static process explicitly, as is discussed in Appendix \ref{app}.
Note that the contributions from the electromagnetic field appearing in the above relation cancel 
and the usual thermodynamic relation is reproduced when the polarization and the magnetization disappear, 
${\bf P} = {\bf M} + {\bf M}_{Pauli}= 0$.

Then the evolution equation of the entropy per unit mass is given by 
\begin{eqnarray}
 T \frac{DS_{en}}{Dt} 
= 
\frac{D U}{Dt} - \frac{D \hat{U}_{em}}{Dt} 
+ 
 \hat{\bf D} \cdot \frac{D^* {\bf E}}{Dt}  
+ \hat{\bf B} \cdot \frac{D^* {\bf H}}{Dt} 
+ (P + U_{em} ) 
\frac{D 1/\rho_m}{Dt} .
\end{eqnarray}
Here we interpret $D/Dt$ as $D^*/Dt$ when it is operated to vectors. 
In fact, there is the following relation,  
\begin{equation}
\frac{D}{Dt} {\bf X}\cdot {\bf Y} = {\bf X} \cdot \frac{D^*}{Dt} {\bf Y} 
+ {\bf X} \cdot \frac{D^*}{Dt} {\bf Y},
\end{equation}
for arbitrary vectors ${\bf X}$ and ${\bf Y}$.

This equation is cast into the following form, 
\begin{equation}
\rho_m \frac{D}{Dt}S_{en} = - \nabla \cdot \frac{\bf q}{T} + \sigma_s,
\end{equation}
where the entropy production is defined by 
\begin{eqnarray}
\sigma_s
&=& 
\left[ 
 \frac{1}{T}{\bf j} \cdot ( {\bf E} +  {\bf v} \times {\bf B} )  \frac{}{} \right. \nonumber \\
&& 
+
\left\{ \tau^D + \frac{\gamma_g \rho_e}{2c} ({\bf s}_m \cdot {\bf B} )  \right\} \frac{1}{T}\theta 
+ \frac{1}{2T} \sum_{jk}(\tau^S)^{jk} e^{jk}  \nonumber \\
&& \left. 
- \frac{1}{T} {\bf p}^a_v \cdot \left( {\bf \Omega} - \frac{1}{4} {\bf \omega}_v \right)
+ \sum_{ik} \frac{1}{T} \tilde{C}^{ik} \partial_k \Omega^{i}
+ {\bf q}\cdot \nabla \frac{1}{T} \right]. \label{entropro}
\end{eqnarray}
The unknown currents and tensors are determined by employing the algebraic positivity of 
this entropy production \cite{degroot}. 
Then  we obtain 
\begin{subequations}
\begin{eqnarray}
&& {\bf j} 
= \sigma ({\bf E} + {\bf v} \times {\bf B}) , \\
&& \tau^D 
= 
\zeta \theta 
- \frac{\gamma_g \rho_e}{2c} {\bf s}_m \cdot {\bf B}, \\
&& (\tau^S)^{ik} 
= 
\eta e^{ik} 
, \\
&& \tilde{C}^{ij}
= 
D_s \partial_j s^{i}_m , \\
&& {\bf q} = \kappa \nabla \frac{1}{T}, \\
&& {\bf p}^a_v 
=
- \lambda 
\left( {\bf \Omega} - \frac{1}{4} {\bf \omega}_v \right).
\end{eqnarray}
\end{subequations}
The first equation corresponds to the Ohm law. The electric conductivity, 
the spin diffusion coefficient, the bulk viscosity, the shear viscosity 
and the heat conductivity are given by $\sigma$, $D_s$, $\zeta$, $\eta$ and $\kappa$, respectively.
The last equation represents the contribution from spin and the vorticity of the fluid,  
and its transport coefficient is given by $\lambda$.

From LIT and the Curie (symmetry) principle \cite{degroot}, the most general irreversible flows are 
given by the linear combinations of currents which have the same transformation properties.
For example, both of ${\bf J}$ and ${\bf q}$ are vector currents 
and the most general expressions of these two currents are given by the linear combinations,
\begin{subequations}
\begin{eqnarray}
{\bf q}_{tot} &=&  L_{11} \nabla \frac{1}{T} + L_{12} \frac{{\bf E} + {\bf v} \times {\bf B}}{T}, \\
{\bf j}_{tot} &=&  L_{21} \nabla \frac{1}{T}  + L_{22}\frac{{\bf E} + {\bf v} \times {\bf B}}{T}.
\end{eqnarray}
\end{subequations}
The coupling between the charge and heat currents are known as the Seebeck and Peltier effects. 
The constraint to $L_{12}$ and $L_{21}$ is given by Onsager's reciprocal theorem \cite{degroot}.

We did not consider the separation of the symmetric and anti-symmetric parts of 
$\tilde{C}^{ij}$ as was done for $(p^s)^{ij}$. 
If we consider such a separation, it is, in principle, 
possible to consider the mixtures between $\partial_j s^{i}_m$ and $\partial_j v^{i}$ as was done for ${\bf q}$ and ${\bf j}$.
This mixture is, however, not considered here.
See also the discussion in Sec. \ref{chap:last}.

Finally, the equations of the spin-electromagnetic hydrodynamics are summarized as 
\begin{subequations}
\begin{eqnarray}
\frac{D}{Dt} \rho_m &=& - \rho_m (\nabla \cdot {\bf v}), \\
\rho_m\frac{D}{Dt} v^{i} &=& -\partial_i ( P - \zeta \theta )
+ \sum_j \partial_j (\eta e^{ij})   
+ \frac{\lambda}{2}\left( \left[\nabla \times \frac{{\bf s}_m}{I} \right]^i + \frac{1}{2} \sum_j \partial_j \omega^{ij} \right)
 \nonumber \\
&& + \rho_e [ {\bf E} +  {\bf v} \times {\bf B} ]^{i} 
+ [ {\bf j}_{tot} \times {\bf B} ]^{i}
+ \sum_l \left[ D^l \partial_i E^l + B^l \partial_i H^l \right] - \partial_i U_{em}, \\
%
\rho_m \frac{D}{Dt}  s^{i}_m  
&=&
-\lambda \left[ \frac{{\bf s}_m}{I} - \frac{1}{4} {\bf \omega}_v \right]^i
+  \sum_{j}\partial_j \left\{ \rho_m [{\bf s}_m \times \partial_j {\bf s}_m ]_i + D_s 
\partial_j s^{i}_m 
\right\} \nonumber \\
&& - \left[ \frac{1}{\varepsilon_0 }{\bf D}\times{\bf E} + \frac{1}{\mu_0} {\bf B} \times ({\bf M}+{\bf M}_{Pauli}) \right]^i, \nonumber \\
\label{finalspineq} \\
%
\rho_m \frac{D}{Dt}  U
&=&
-\frac{D}{Dt} U_{em} 
+ 
 {\bf E} \cdot \frac{D^*}{Dt} {\bf D}  
+ {\bf H} \cdot \frac{D^*}{Dt} {\bf B}
+ {\bf j}_{tot}\cdot [{\bf E} + ({\bf v} \times {\bf B})]  
+ P  \frac{1}{\rho_m} \frac{D}{Dt} \rho_m  \nonumber \\
&&
+ \frac{1}{3} \zeta \theta^2  
+ \frac{\eta}{2} \sum_{jk} (e^{jk})^2 
+ \lambda \left( \frac{{\bf s}_m}{I} - \frac{1}{4} {\bf \omega}_v \right)^2
+ \frac{D_s}{I}\sum_{ij} (\partial_j s^{i}_m)^2
- \nabla \cdot {\bf q}_{tot}  . \label{gibbs1}
\end{eqnarray}
\end{subequations}

\section{Equation of state dependence}

In this section, we consider a special case where the 
the properties of the polarization and the magnetization are restricted as follows,
\begin{subequations}
\begin{eqnarray}
{\bf P} &=& \chi_E {\bf E}, \\
{\bf M} + {\bf M}_{Pauli} &=& \chi_H {\bf H},
\end{eqnarray}
\end{subequations}
where $\chi_E$ and $\chi_H$ are the electric and magnetic susceptibilities, respectively.
That is, ${\bf D}$ ($\bf {B}$) is proportional to ${\bf E}$ (${\bf H}$).
Then the thermodynamic relation (\ref{eos1}) is reduced to 
\begin{eqnarray}
TdS_{en} 
&=&
d U  
+
(P + U_{em}) 
d \frac{1}{\rho_m} 
- \frac{1}{2}
 \left[ 
 {\bf E}^2 d \frac{\chi_E}{\rho_m} 
+ {\bf H}^2  d \frac{\chi_H}{\rho_m}  \right] \label{eos2}
.
\end{eqnarray}

So far, we have used the thermodynamic relation among 
$dU$, $dS_{en}$, $d\rho_m$, $d(\chi_E/\rho_m)$ and $d(\chi_H/\rho_m)$.
For the later convenience, we choose $\chi_E$ and $\chi_H$ to be thermodynamic variables instead of 
$\chi_E/\rho_m$ and $\chi_H/\rho_m$, respectively.
Then the thermodynamic relation (\ref{gibbs1}) is, again, re-expressed as
\begin{eqnarray}
TdS_{en} 
&=&
d U  
+
P
d \frac{1}{\rho_m} 
- \frac{1}{2\rho_m}
 \left[ 
 {\bf E}^2 d \chi_E 
+ {\bf H}^2 d \chi_H  \right].
\end{eqnarray}

Following Refs. \cite{degroot,chu,imai}, we consider the free energy $F = U- TS_{en}$ and its variation, 
\begin{eqnarray}
dF 
&=&
d(U - TS_{en}) \nonumber \\
&=&
 - \left( P + \rho_m \frac{{\bf E}^2}{2}\frac{\partial \chi_E}{\partial \rho_m} 
+ \rho_m \frac{{\bf H}^2}{2}\frac{\partial \chi_H}{\partial \rho_m} \right)
d\frac{1}{\rho_m} 
- \left( S_{en} - \frac{{\bf E}^2}{2\rho_m}\frac{\partial \chi_E}{\partial T} 
- \frac{{\bf H}^2}{2\rho_m}\frac{\partial \chi_H}{\partial T} \right) dT . \nonumber \\
\end{eqnarray}
Here we assumed that $\chi_E$ and $\chi_H$ are functions of $\rho_m$ and $T$.
Thus we can eliminate the $\chi_E$ and $\chi_H$ dependences by 
introducing the following new pressure and entropy which are functions only of $\rho_m$ and $T$,
\begin{eqnarray}
dF = - P_0 d\frac{1}{\rho_m} - S_0 dT ,
\end{eqnarray}
where
\begin{subequations}
\begin{eqnarray}
P &=& P_0 (\rho_m,T) -  \rho_m \frac{{\bf E}^2}{2}\frac{\partial \chi_E}{\partial \rho_m} 
- \rho_m \frac{{\bf H}^2}{2}\frac{\partial \chi_H}{\partial \rho_m}, \\
S_{en} &=& S_0 (\rho_m,T) + \frac{{\bf E}^2}{2\rho_m}\frac{\partial \chi_E}{\partial T} 
+ \frac{{\bf H}^2}{2\rho_m}\frac{\partial \chi_H}{\partial T} .
\end{eqnarray}
\end{subequations}

By using this new pressure, our momentum equation is re-expressed as 
\begin{eqnarray}
\rho_m \frac{D}{Dt} {\bf v}^{i} 
&=& -\partial_i (P_0 - \zeta \theta ) 
+ \sum_j \partial_j (\eta e^{ij}) + \rho_e [{\bf E} +  {\bf v} \times {\bf B} ]^{i} + [ {\bf j}_{tot} \times {\bf B} ]^{i} \nonumber \\
&& - \frac{{\bf E}^2}{2}\partial_i \chi_E 
- \frac{{\bf H}^2}{2}\partial_i \chi_H 
+ \partial_i \left( \rho_m \frac{{\bf E}^2}{2}\frac{\partial \chi_E}{\partial \rho_m} \right)
+ \partial_i \left( \rho_m \frac{{\bf H}^2}{2}\frac{\partial \chi_H}{\partial \rho_m} \right).
\end{eqnarray}
Here, for the sake of comparison, we ignore the contributions from spin.
The second line on the right hand side is equivalent to the so-called Korteweg-Helmholtz force, which is 
known as one of the possible candidates for the interactions between fluid and the electromagnetic field \cite{bobbio}.
The Kelvin force is another candidate and 
the difference between the Kelvin and Korteweg-Helmholtz forces is absorbed into the 
definition of the pressure again. See the discussion in Refs. \cite{degroot,bobbio}.

\section{Dynamics of polarization and magnetization} \label{chp:pola}

We have considered that the electromagnetic field is 
the solution of the Maxwell equation for macroscopic medium. 
On the other hand, as is discussed in Refs. \cite{degroot,eu-oppen,fk}, it is also possible to construct constitutive equations of the polarization 
and the magnetization 
by introducing the two different electromagnetic fields: one is the solution of the Maxwell equation for macroscopic medium and the other the 
conjugate variables of the polarization and the magnetization appearing in the thermodynamic relation. 
The latter fields are denoted by ${\bf E}^{eq}$ and ${\bf H}^{eq}$, and then the thermodynamic relation is re-expressed as
\begin{equation}
dU + \frac{1}{\rho_m}d \frac{1}{2}({\bf E} \cdot {\bf P} 
+ {\bf H} \cdot {\bf M}_{tot}) 
= 
\hat{\bf E}^{eq} \cdot d {\bf P} + \hat{\bf H}^{eq} \cdot d {\bf M}_{tot}
-P d \frac{1}{\rho_m} + T dS_{en},
\end{equation}
where ${\bf M}_{tot} = {\bf M} + {\bf M}_{Pauli}$.

By using this relation, we can calculate the entropy production again and obtain
\begin{eqnarray}
\sigma_s 
&=& 
\left[ 
 \frac{1}{T}{\bf j} \cdot ( {\bf E} +  {\bf v} \times {\bf B} )  \frac{}{} \right. \nonumber \\
&& 
+
\left\{ \tau^D + \frac{\gamma_g \rho_e}{2c} ({\bf s}_m \cdot {\bf B} )  \right\} \frac{1}{T}\theta 
+ \frac{1}{2T} \sum_{ik} (\tau^S)^{jk} e^{jk} \nonumber \\
&& \left. 
- \frac{1}{T} {\bf p}^a_v \cdot \left( {\bf \Omega} - \frac{1}{4} {\bf \omega}_v \right)
+ \sum_{ik} \frac{1}{T} \tilde{C}^{ik} \partial_k \Omega^{i}
+ {\bf q}\cdot \nabla \frac{1}{T} \right. \nonumber \\
&& \left. + \frac{1}{T}({\bf E} - {\bf E}^{eq}) \frac{D^*}{Dt} {\bf P} 
+ \frac{1}{T}({\bf H} - {\bf H}^{eq}) \frac{D^*}{Dt} {\bf M}_{tot}
\right]. \nonumber \\
\end{eqnarray}

The new contribution to Eq. (\ref{entropro}) 
comes from the last two terms on the right hand side of the equation. 
The positivity of the entropy production for these terms then leads to 
\begin{subequations}
\begin{eqnarray}
\frac{D}{Dt} {\bf P} - {\bf \Omega} \times {\bf P} 
&=& \frac{1}{\tau_p} ({\bf E} - {\bf E}^{eq}), \\
\frac{D}{Dt} {\bf M}_{tot} - {\bf \Omega} \times {\bf M}_{tot} 
&=& \frac{1}{\tau_M} ({\bf H} - {\bf H}^{eq} ),
\end{eqnarray}
\end{subequations}
where $\tau_p$ and $\tau_m$ are new transport coefficients, characterizing the time scales of the relaxations.

There are several proposals for the evolution equations of the polarization and the magnetization.
In Refs. \cite{degroot,eu-oppen}, the derived equations of the polarization and the magnetization do not contain the terms 
representing the rotation of the polarization and the magnetization, which corresponds to 
the second terms on the left hand side of the above equations. 
This is because the internal angular momentum or spin is not considered as a hydrodynamic variable.

The same rotation terms are obtained in Ref. \cite{fk} where   
the rotational vector ${\bf \Omega}$ is replaced by $(\nabla \times {\bf v})/2$, and 
it is discussed that 
the derived equation of the polarization is close to Eq. (2.6) of Ref. \cite{ho}. 
If we can use the same replacement in our result, 
one can see that our equation is more similar to Eq. (2.6) of Ref. \cite{ho} compared to that of Ref. \cite{fk}.

\section{Comparison with other theories}

As was mentioned in the introduction, there are three different hydrodynamic models 
including the spin degree of freedom (internal angular momentum) and the electromagnetic field. 
In this section, we would like to summarize the comparison among them.

In Ref. \cite{mark}, the spin-magnetohydrodynamics is obtained by using the result 
by Bohm et al.
As was mentioned in the introduction, however, the equation of the energy density is not obtained.
Moreover, the form of the stress tensor, which corresponds to viscosity, 
is not calculated in the framework of this theory.

In Ref. \cite{fk}, the model is obtained phenomenologically using the definition of de Groot-Mazur. 
In this derivation, the positivity of the entropy production is satisfied in the last of the derivation using LIT. 
However, the form of the fluid stress tensor is assumed to be given by Eq. (4.7) 
of Ref. \cite{fk} and the approach of LIT is not completely employed.
Because of this incompleteness, the mixing effect of irreversible currents such as the Seebeck and Peltier effect 
is not discussed. 
Furthermore, the internal angular momentum included in this theory is not equivalent to spin. 
For example, the traditional conservation equation is assumed for the internal angular momentum, 
\begin{equation}
\partial_t {\bf s}_v + \nabla \cdot ({\bf s}_v {\bf v}) = {\rm source~terms},
\end{equation}
where ${\bf s}_v$ is the internal angular momentum density of Ref. \cite{fk}.
See Eq. (4.3) of Ref. \cite{fk}.
However, this structure is different from our spin equation (\ref{finalspineq}), which is the non-linear equation of ${\bf s}_m$.
Thus it is not adequate to identify the internal angular momentum discussed in Ref. \cite{fk} with spin.
Moreover, as was shown in the previous section, 
our polarization equation is more promising than that of Ref. \cite{fk}.

Our approach is the extension of the argument of Ref. \cite{imai} 
to the case with the spin degree of freedom as was mentioned in the introduction. 
Thus same thermodynamic relation and electromagnetism for macroscopic medium are used.
However, the internal angular momentum introduced in this work is essential same as that of Ref. \cite{fk} and hence 
is different from our spin equation.
Moreover, the mixing effect of the irreversible flows and the dynamics of the polarization and the magnetization are not investigated.

If the spin degree of freedom (or the internal angular momentum) is neglected, sEMHD is reduced to electromagnetic hydrodynamics. 
For the purpose of reference, we will comment some results of this type of theory.

The result of Ref. \cite{degroot} is the essentially same as Ref. \cite{fk} except for the absence of the internal angular momentum.
However, differently from Ref. \cite{fk}, the fluid stress tensor 
is determined by employing the positivity of the entropy production completely as was done in this paper. 
Moreover, the mixing effect of the irreversible currents is also taken into account.

In Ref. \cite{eu-oppen}, the derivations of electromagnetic hydrodynamics are considered for 
three different choices of electromagnetism for macroscopic medium: the Minkowski, Hertz-Abraham and de Groot-Mazur definitions.
It is concluded that the choice of Minkowski is promising, but 
the derived entropy production does not satisfy the algebraic positivity and 
the form of the viscous tensor and the heat current are not evaluated.
The thermodynamic relation used in Ref. \cite{eu-oppen} (given by Eq. (4.17)) is different from what we used in this work
and this is the reason why the positivity of the entropy production is successfully satisfied in our derivation.

The thermodynamic relation used in Ref. \cite{chu} is the same as ours (See Eq. (5.10b) in Ref. \cite{chu}), 
while the Abraham-Becker stress tensor is used for electromagnetism for macroscopic medium. 
Moreover, the form of the fluid stress tensor is assumed 
and is not obtained by using the positivity of the entropy production.

\section{Concluding remarks} \label{chap:last}

The hydrodynamic model including the spin degree of freedom and the electromagnetic field was derived.
In this derivation, we applied electromagnetism for macroscopic medium proposed by Minkowski.
For the evolution equation of spin, we assumed that the hydrodynamic equation of the Pauli equation 
is reproduced when the many-body effect is neglected. 
The many-body effect for the fluid and spin stress tensors 
were obtained by employing the algebraic positivity of the entropy production 
in the linear irreversible thermodynamics.



As was shown in this paper, the positivity 
of the entropy production is satisfied in the derivation of sEMHD even if we use the definition of Minkowski.
This conclusion is different from that of Ref. \cite{fk}, 
where it is claimed that the definition of de Groot-Mazur should be employed for the consistent theory.
In fact, our derivation is consistent with LIT more than Ref. \cite{fk}, 
because the assumed form of the stress tensor is not employed in our approach. 
Furthermore, our polarization equation can reproduce the result of Ref. \cite{ho} better than 
Ref. \cite{fk}.

However, it does not necessarily mean that we should choose the Minkowski definition, because 
it may be still possible to derive a consistent hydrodynamic model in the de Groot-Mazur definition 
by considering the different thermodynamic relation. 
Similarly, it is also important to confirm the applicability of the definition by Hertz-Abraham to sEMHD.
The superiority of these derived hydrodynamic models will be studied by, for example, 
the stability analysis of the models as was done for the relativistic hydrodynamic models \cite{stabili}.

In the present argument, we discussed the energy conservation by considering the spin vector 
per unit mass as an analogue of the classical angular momentum.
Then the spin rotation energy is assumed to be given by $I {\bf \Omega}^2$. 
In fact, it is known that the kinetic energy of a quantum particle with spin can be expressed by 
the sum of the kinetic term of velocity and the rotation of spin, as is shown in Ref. \cite{hest}, 
\begin{equation}
\frac{1}{2m}\langle {\bf p}^2 \rangle + \frac{{\bf s}^2}{2I}, \label{qua_mi}
\end{equation}
where 
\begin{equation}
I = \frac{m}{\langle (\nabla \ln \psi^* \psi)^2 \rangle}.
\end{equation}
Here $\langle~\rangle$ represents the quantum mechanical expectation value, respectively.
One can see that this $I$ is the essentially same as the classical moment of inertia.
In fact, when the Gaussian distribution is considered, $\psi^* \psi \approx e^{-{\bf r}^2/(2\sigma^2)}$, we obtain $I = m\sigma^2$. 
Thus our classical treatment of the spin degree of freedom can be justified 
when the fluid is given by the ensemble of wave packets of constituent particles of a fluid with a well-defined width $\sigma$.

We developed the model of sEMHD by taking notice of the classical 
aspect of the spin degree of freedom.
Then the many-body behaviors of the original quantum dynamics is not reproduced exactly. 
For example, the effect of the Pauli exclusion principle will not be included. 
Thus our model is applicable for high temperature systems where 
this effect becomes relatively small.

We considered the mixture of the two vector currents and obtained 
the Seebeck and Peltier effects.
In principle, it is also possible to consider the similar mixture for tensors.
For example, the shear stress tensor can be modified by the mixture with the spin tensor as 
\begin{eqnarray}
\eta e^{ij} \longrightarrow L^{sv}_{11} e^{ij} + L^{sv}_{12} e^{ij}_s ,
\end{eqnarray}
with 
\begin{equation}
e^{ij}_s = \partial_j s^i_m - \partial_i s^j_m .
\end{equation}
The magnitude of the mixture is determined by Onsager's reciprocal theorem.

In this discussion, we did not consider the multi-component fluids with different charges. 
The generalization of our approach to such a case will be implemented 
following the argument of Ref. \cite{degroot}.

As was discussed in the introduction, 
we would like to construct the relativistic hydrodynamic model including the effect of spin, but the 
generalization of the present argument to the relativistic case is not trivial, 
because spin is not an appropriate quantum number for relativistic systems.
However, the consistent relativistic theory should reproduce our result in the non-relativistic limit, 
as the relativistic hydrodynamic model reproduces the Navier-Stokes equation in the appropriate non-relativistic limit.
As one of examples of relativistic hydrodynamics associated with spin, see Ref. \cite{becattini}.

\vspace{1cm}

The author acknowledges C. E. Aguir, T. Hirano, X. Huang and T. Kodama the members of the nuclear theory groups of Sophia university and Riken 
for useful discussions and comments.
This work was financially supported by CNPq.

\appendix

\section{Thermodynamic Relation} \label{app}

In this appendix, we derive the thermodynamic relation given by Eq. (\ref{eos1}).

The general expression of the first law of thermodynamics is 
\begin{equation}
\Delta Q = dE + dW, 
\end{equation}
where $\Delta Q$ and $dW$ represent heat and the thermodynamic work, respectively.
These quantities are expressed by thermodynamic variables in the quasi-static process.
In particular, the usual thermodynamic work in this process is expressed as
\begin{equation}
dW = PdV.
\end{equation}

When we have the electromagnetic field, it acts as an external field to a thermodynamic system and 
gives rise to another contribution to the above thermodynamic work \cite{reichl}.
The contribution is estimated as follows.

We separate a fluid into the ensemble of fluid elements (cells) and assume that the thermal equilibrium is 
satisfied inside of these elements. 
On the other hand, the change of the momentum of the fluid is induced by the stress tensors acting on the surface of the fluid element.
Then the change of the work for a fluid element is defined by 
\begin{eqnarray}
\frac{D}{Dt}W 
&=& \sum_{ik}\int dS v^{i}(p^{ik}+T^{ik})n^k .
\end{eqnarray}
This quantity is calculated as 
\begin{eqnarray}
\frac{D}{Dt}W 
&=& \int dV \left[ {\bf j}\cdot ({\bf E}+{\bf v} \times {\bf B}) + \rho_m {\bf E} \frac{D}{Dt} \hat{\bf D} 
+ \rho_m  {\bf H} \frac{D}{Dt} \hat{\bf B}\right]\nonumber \\
&& + \int dV \left[ \frac{\rho_m }{2} \frac{D}{Dt}{\bf v}^2 + \sum_{ij} p^{ij} \partial_j v^{i} - U_{em} \nabla \cdot {\bf v} \right].
\end{eqnarray}

To obtain the correction to the thermodynamic work $PdV$ (or $Pd(1/\rho_m)$ in the notation of hydrodynamics), 
we consider the quasi-static process which is reversible.
Then we ignore the contribution corresponding to the higher order of the fluid velocity, ${\bf v}^2 = 0$ and 
the contribution from the irreversible flows.
The thermodynamic work in the quasi-static process is finally obtained by 
\begin{eqnarray}
dW 
&=& \int dV \rho_m \left[ 
- (P + U_{em}) d \frac{1}{\rho_m} + {\bf E} d \hat{\bf D} 
+ {\bf H} d \hat{\bf B}
\right]. 
\end{eqnarray}
Thus one can see that, because of the spin (the magnetization) and the electromagnetic degrees of freedom, 
the thermodynamic work is modified as
\begin{equation}
P d\frac{1}{\rho_m} \longrightarrow 
(P + U_{em}) d \frac{1}{\rho_m} - {\bf E} d \hat{\bf D} 
- {\bf H} d \hat{\bf B}.
\end{equation}
As a result, the thermodynamic relation is modified as Eq. (\ref{eos1}) in the present case.

The similar result is obtained in Ref. \cite{chu}.

\end{document}